\documentclass[prd,twocolumn,superscriptaddress,amssymb,amsfonts,amsmath,nofootinbib,preprintnumbers,aps]{revtex4-2}
\usepackage{graphicx,slashed,xcolor,multirow,bbold,mathtools,sidecap,tikz,bm,ulem,enumitem,booktabs,array,inputenc}
\usepackage[colorlinks=true,linktoc=page,linkcolor=purple,citecolor=teal,urlcolor=magenta]{hyperref}
\usepackage[compat=1.1.0]{tikz-feynman}
\graphicspath{ {F95/} }
\usepackage{siunitx,adjustbox,comment}
\definecolor{lime}{HTML}{A6CE39}
\DeclareRobustCommand{\orcidicon}{\hspace{-2.1mm}
\begin{tikzpicture}
\draw[lime,fill=lime] (0,0.0) circle [radius=0.13] node[white] {{\fontfamily{qag}\selectfont \tiny ID}}; \draw[white,fill=white] (-0.0525,0.095) circle [radius=0.007]; 
\end{tikzpicture} \hspace{-3.7mm} }
\foreach \x in {SA,GC,AC,SM,AT} {\expandafter\xdef\csname orcid\x\endcsname{\noexpand\href{https://orcid.org/\csname orcidauthor\x\endcsname} {\noexpand\orcidicon}}}

\let\emph\textit

\begin{document}

\preprint{TTP25-005, P3H-25-013, ZU-TH 12/25, ICPP-91}

\title{LHC Signatures of the Generic Georgi-Machacek Model}

\author{Saiyad Ashanujjaman\orcidSA{}}
\email{saiyad.ashanujjaman@kit.edu}
\affiliation{Institut f\"ur Theoretische Teilchenphysik, Karlsruhe Institute of Technology, Engesserstra\ss e 7, D-76128 Karlsruhe, Germany}
\affiliation{Institut f\"ur Astroteilchenphysik, Karlsruhe Institute of Technology, Hermann-von-Helmholtz-Platz 1, D-76344 Eggenstein-Leopoldshafen, Germany}

\author{Andreas Crivellin\orcidAC{}}
\email{andreas.crivellin@cern.ch}
\affiliation{Physik-Institut, Universität Zürich, Winterthurerstrasse 190, CH–8057 Zürich, Switzerland}

\author{Siddharth P.~Maharathy\orcidSM{}}
\email{siddharth.prasad.maharathy@cern.ch}
\affiliation{School of Physics and Institute for Collider Particle Physics, University of the Witwatersrand, Johannesburg, Wits 2050, South Africa}
\affiliation{Indian Institute of Science Education and Research Pune, Dr.~Homi Bhabha Road, Pune 411008, India}

\author{Anil Thapa\orcidAT{}}
\email{a.thapa@colostate.edu}
\affiliation{Physics Department, Colorado State University, Fort Collins, CO 80523, USA}

\begin{abstract}
Vector-boson fusion production of new Higgs bosons decaying into pairs of electroweak gauge bosons ($W^\pm W^\pm$, $WZ$ and $ZZ$) is a smoking-gun signature of the Georgi-Machacek (GM) Model. Notably, ATLAS has observed a $3.3\sigma$ excess in $W^\pm W^\pm$ at $\approx 450$\,GeV and a $2.8\sigma$ excess in the $WZ$ channel at $\approx 375$\,GeV, while CMS reported weaker-than-expected limits at these masses. However, the canonical custodial-symmetric GM Model cannot accommodate these signals, as it predicts mass degeneracy among the new gauge-philic Higgs bosons. To overcome this obstacle, we consider a generalized version of the GM Model without the custodial $SU(2)_C$ symmetry in the scalar potential. In the limit of small mixing among the Higgs bosons, the $W^\pm W^\pm$ and $WZ$  excesses can be explained by the doubly and singly-charged Higgs bosons originating primarily from the $Y=1$ triplet, while respecting the bounds from $ZZ$ searches. Furthermore, the neutral Higgs boson mostly contained in the $Y=0$ triplet can account for the excess at $\approx 152$\,GeV in associated di-photon production, while being consistent with constraints from vacuum stability and the Standard Model Higgs signal strength measurements. 
\end{abstract}

\maketitle

\section{Introduction} 
\label{sec:intro}
The Standard Model (SM) of particle physics is the currently accepted theoretical description of the known fundamental constituents and interactions of matter. It has been successfully tested experimentally~\cite{ParticleDataGroup:2022pth}, culminating in the discovery of the Higgs boson at the Large Hadron Collider (LHC)~\cite{Aad:2012tfa, Chatrchyan:2012ufa}. Although the observed $125$\,GeV scalar has properties consistent with those predicted by the SM~\cite{ATLAS:2022vkf,CMS:2022dwd}, this does not rule out the existence of additional scalar bosons. In fact, as long as their contribution to electroweak symmetry (EW) breaking is subleading and they approximately respect custodial symmetry ($\rho=m_W^2/(m_Z^2 c_W^2)\approx 1$, where $c_W$ is the cosine of the Weinberg angle), they are in general compatible with the bounds from SM Higgs signal strength measurements~\cite{ATLAS:2022vkf,CMS:2022dwd} and electroweak (EW) precision data~\cite{Haller:2022eyb,deBlas:2021wap}.

While the vacuum expectation values (VEVs) of $SU(2)_L$ singlets and doublets preserve $\rho=1$ at tree level, higher $SU(2)_L$ representations do not. Therefore, if one supplements the SM with an $SU(2)_L$ triplet of hypercharge $Y=1$ or $Y=0$, its VEV must be small---of the order of a few GeV---to result in a $W$ mass prediction compatible with experiments. Because the couplings of the scalars to a pair of EW gauge bosons are proportional to their VEVs, their vector boson fusion (VBF) cross sections, i.e.~the production in association with forward jets, are also small. Consequently, despite the distinct experimental signatures, VBF processes in these models are beyond the LHC reach. 

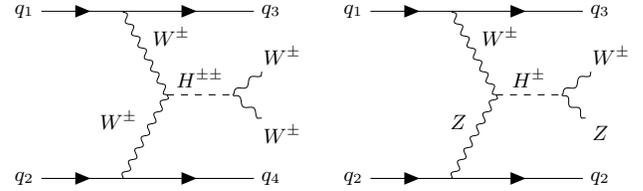
\begin{figure}[htb!]
\resizebox{0.99\columnwidth}{!}{
\begin{tikzpicture}
\begin{feynman}
\vertex(a);
\vertex[above right=1.5cm of a] (b){$q_1$};
\vertex[right=1.5cm of b] (b1);
\vertex[right=2cm of b1] (b2){$q_3$};
\vertex[below right=1.5cm of a] (c){$q_2$};
\vertex[right=1.5cm of c] (c1);
\vertex[right=2cm of c1] (c2){$q_4$};
\vertex[right=3.5cm of a] (a1);
\vertex[right=1cm of a1] (a2);
\vertex[above right=0.5cm of a2] (d1){$W^\pm$};
\vertex[below right=0.5cm of a2] (d2){$W^\pm$};
\diagram* {
(b)-- [fermion] (b1),
(b1)-- [fermion] (b2),
(c)-- [fermion] (c1),
(c1)-- [fermion] (c2),(b1)-- [boson, edge label=$W^\pm$] (a1),(c1)-- [boson, edge label=$W^\pm$] (a1),(a1)-- [scalar, edge label=$H^{\pm\pm}$]  (a2),(a2)-- [boson] (d1),(a2)-- [boson] (d2);};
\end{feynman}
\end{tikzpicture}
\quad
\begin{tikzpicture}
\begin{feynman}
\vertex(a);
\vertex[above right=1.5cm of a] (b){$q_1$};
\vertex[right=1.5cm of b] (b1);
\vertex[right=2cm of b1] (b2){$q_3$};
\vertex[below right=1.5cm of a] (c){$q_2$};
\vertex[right=1.5cm of c] (c1);
\vertex[right=2cm of c1] (c2){${q_2}$};
\vertex[right=3.5cm of a] (a1);
\vertex[right=1cm of a1] (a2);
\vertex[above right=0.5cm of a2] (d1){$W^\pm$};
\vertex[below right=0.5cm of a2] (d2){$Z$};
\diagram* {
(b)-- [fermion] (b1),
(b1)-- [fermion] (b2),
(c)-- [fermion] (c1),
(c1)-- [fermion] (c2),(b1)-- [boson, edge label=$W^\pm$] (a1),(c1)-- [boson, edge label=$Z$] (a1),(a1)-- [scalar, edge label=$H^{\pm}$]  (a2),(a2)-- [boson] (d1),(a2)-- [boson] (d2);};
\end{feynman}
\end{tikzpicture}
}
\caption{Feynman diagrams showing the VBF production of singly and doubly charged Higgses decaying to EW gauge bosons.}
\label{FeynmanDiagrams}
\end{figure}

\begin{figure*}[t!]
\centering
\includegraphics[scale=0.5]{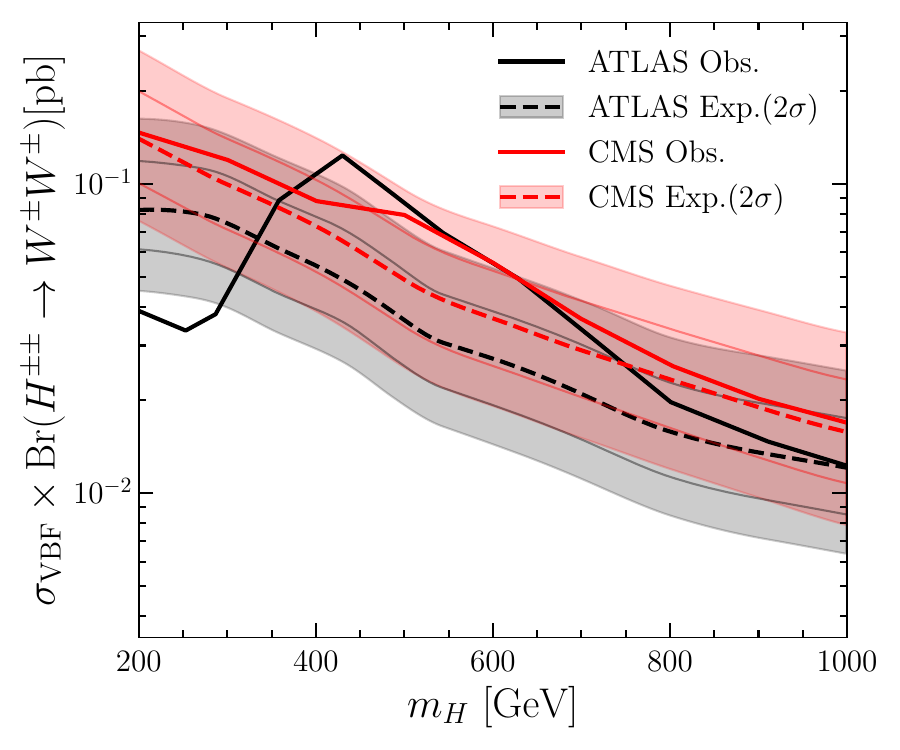}
\includegraphics[scale=0.5]{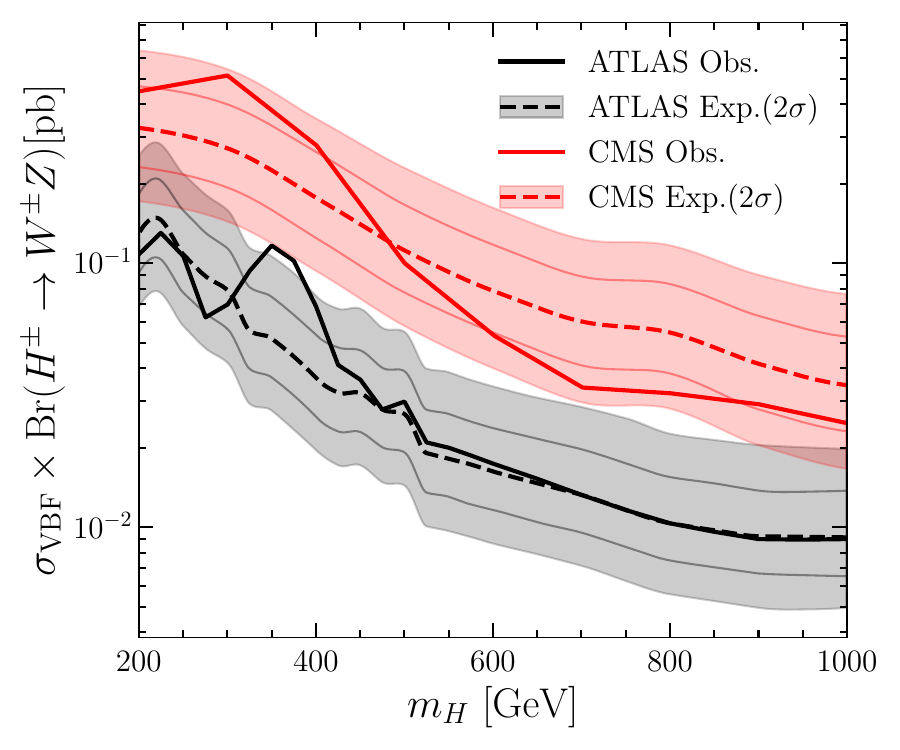}
\caption{Observed and expected limits on VBF production of new Higgses decaying to $W^\pm W^\pm$ and $WZ$ by ATLAS~\cite{ATLAS:2022zuc,ATLAS:2023sua} and CMS~\cite{CMS:2021wlt}, based on the full run-2 LHC data set. Note that the excesses in the ATLAS analyses, which are most pronounced at $450$\,GeV ($3.3\sigma$) and $375$\,GeV ($2.8\sigma$), overlap with the regions where CMS observes weaker-than-expected limits.}
\label{fig:ZWWWexp}
\end{figure*}

Georgi and Machacek (GM) proposed supplementing the SM with a $Y=0$ and a $Y=1$ scalar triplet. By imposing custodial symmetry, the triplets are arranged into an $SU(2)_C$ ($C\equiv$ custodial) multiplet~\cite{Georgi:1985nv,Chanowitz:1985ug} so that $\rho=1$ at the tree level. This allows sizable triplet VEVs, leading to potentially measurable VBF production cross sections for new Higgses at the LHC. In particular, VBF production of $W^\pm W^\pm$ and $WZ$ are smoking-gun signatures of the GM model~\cite{Gunion:1989ci,Godfrey:2010qb,Chiang:2012cn,Englert:2013zpa,Hartling:2014zca,Godunov:2014waa,Logan:2015xpa,Chiang:2015kka,Chiang:2015amq,Chen:2023bqr,Ghosh:2023izq,Chakraborti:2023mya,Ahriche:2022aoj,Ghosh:2022bvz,deLima:2022yvn,Chen:2022zsh,Chen:2022ocr,Wang:2022okq,Zhu:2020bew,Ismail:2020kqz,Das:2018vkv,Yang:2018pzt,Keeshan:2018ypw,Chiang:2018cgb,Logan:2017jpr,Chang:2017niy} (see Fig.~\ref{FeynmanDiagrams}), as they are absent in two-Higgs-doublet models (or loop-suppressed for $WZ$ in the case of CP-violation~\cite{Kanemura:2024ezz}). Interestingly, ATLAS observed excesses in the $W^\pm W^\pm$~\cite{ATLAS:2023sua} and $WZ$ channels~\cite{ATLAS:2022zuc}, with local significances of $3.3\sigma$ at 450\,GeV and $2.8\sigma$ at 375\,GeV, respectively. Furthermore, CMS reported weaker-than-expected limits in these mass regions~\cite{CMS:2021wlt} (see Fig.~\ref{fig:ZWWWexp}). Combining these data, Ref.~\cite{LeYaouanc:2023zvi} finds a combined global significance of $4\sigma$ for the $W^\pm W^\pm$ and $WZ$ excesses.

However, these indications for new Higgs bosons cannot be explained within the canonical custodial-symmetric GM model, as it predicts the gauge-philic Higgs bosons to be mass-degenerate. To overcome this limitation, we consider the generic GM model (gGMM)~\cite{Blasi:2017xmc,Keeshan:2018ypw,Chen:2023ins}, where the requirement of custodial symmetry in the Higgs potential is relaxed. In this framework, the mass degeneracy among the Higgses is lifted, while the VEVs of the two triplet Higgses have to remain approximately equal in size, as required by EW precision data. Furthermore, we examine whether the indications for a new Higgs in associated di-photon production at $152$\,GeV~\cite{Crivellin:2021ubm,Bhattacharya:2023lmu,Ashanujjaman:2024pky,Crivellin:2024uhc,Banik:2024ugs,Ashanujjaman:2024lnr} obtained from the side-bands of SM Higgs analyses~\cite{ATLAS:2023omk,ATLAS:2024lhu}\footnote{The existence of a new scalar boson with a mass of around 150\,GeV is a prediction~\cite{vonBuddenbrock:2017gvy,Buddenbrock:2019tua,Coloretti:2023wng,Banik:2023vxa} of the multi-lepton anomalies~\cite{Fischer:2021sqw, Crivellin:2023zui}.} can be explained within the gGMM.\footnote{An explanation of these excesses within the GM model, albeit without detailed calculation, was proposed in Ref.~\cite{Richard:2021ovc}.}

\section{The Model}
\label{sec:model}
Like the canonical GM model, the gGMM is constructed by adding an $SU(2)_L$ triplet scalar $\chi$ with $Y=1$ and a triplet scalar $\zeta$ with $Y=0$ to the SM $SU(2)_L$ Higgs doublet $\phi$. These Higgses decompose into their $SU(2)_L$ components as
\begin{align*}
\phi = \begin{pmatrix} \phi^+ \\ \phi^0 \end{pmatrix},~
\chi = \begin{pmatrix} \frac{\chi^+}{\sqrt{2}} & -\chi^{++} \\ \chi^0 & -\frac{\chi^+}{\sqrt{2}} \end{pmatrix},~
\zeta = \begin{pmatrix} \frac{\zeta^0}{\sqrt{2}} & -\zeta^{+} \\ -\zeta^- & -\frac{\zeta^0}{\sqrt{2}} \end{pmatrix},
\end{align*}
with
\begin{align*}
\phi^0 = \frac{\phi_r + i \phi_i + v_\phi}{\sqrt{2}},~
\chi^0 = \frac{\chi_r + i \chi_i}{\sqrt{2}} + v_\chi,~
\zeta^0 = \zeta_r + v_\zeta,
\end{align*}
where $v_\phi$, $v_\chi$ and $v_\zeta$ are the respective VEVs. With these conventions, the $\rho$ parameter at tree level is given by 
\begin{align}
\rho = \frac{v^2}{v^2+4(v_\chi^2 - v_\zeta^2)},
\label{eq:rho}
\end{align}
where $v^2 = v_\phi^2 + 4(v_\chi^2 + v_\zeta^2) \approx (246\,{\rm GeV})^2$. From $\rho\approx 1$, it follows that $v_\chi\approx v_\zeta$. Furthermore, we know experimentally that the 125\,GeV Higgs is SM-like, i.e.~$h\approx \phi_r$, which implies $v\approx v_\phi$ or $v_\chi^2,v_\zeta^2 \ll v^2$.

Without the custodial symmetry of the canonical GM model, the scalar potential is given by~\cite{Chen:2023ins}
\begin{align}
V = &-m_\phi^2(\phi^\dagger\phi) - m_\chi^2 {\rm Tr}(\chi^\dagger \chi) - m_\zeta^2 {\rm Tr}(\zeta^2) \nonumber
\\
&+(\mu_{\phi\chi}\phi^\dagger\chi\tilde{\phi} + {\rm h.c.}) + \mu_{\phi\zeta}\phi^\dagger \zeta \phi+\mu_{\chi\zeta}{\rm Tr} (\chi^\dagger\chi\zeta) \nonumber
\\
&+\lambda(\phi^\dagger\phi)^2 + \rho_1 [{\rm Tr}(\chi^\dagger\chi)]^2  + \rho_2 {\rm Tr}(\chi^\dagger\chi \chi^\dagger\chi) \nonumber
\\
&+ \rho_3 [{\rm Tr}(\zeta^2)]^2 + \rho_4 {\rm Tr}(\chi^\dagger\chi){\rm Tr}(\zeta^2)+\rho_5 {\rm Tr}(\chi^\dagger\zeta){\rm Tr}(\zeta\chi) \nonumber
\\
&+\sigma_1 {\rm Tr}(\chi^\dagger\chi)\phi^\dagger\phi + \sigma_2 \phi^\dagger\chi\chi^\dagger\phi+\sigma_3 {\rm Tr}(\zeta^2)\phi^\dagger\phi \nonumber
\\
&+(\sigma_4\phi^\dagger\chi\zeta\tilde{\phi}+{\rm h.c.}),
\label{eq:pot}
\end{align}
where only $\mu_{\phi\chi}$ and $\sigma_4$ can be complex (their phases are related~\cite{Chen:2023ins}), 
but are assumed to be real, albeit for simplicity, while the other parameters are real by construction.

As outlined in the introduction, we aim at a proof of principle that the $W^\pm W^\pm$ and $WZ$ excesses can be explained in the gGMM while remaining consistent with other bounds and accommodating the indication for a 152\,GeV di-photon resonance. For this purpose, we consider a simplified setup in which the $Y=1$ and $Y=0$ triplets, $\chi$ and $\zeta$, are decoupled from each other in a first approximation. This is achieved by setting $\mu_{\chi\zeta} \approx 0, \rho_4 \approx 0, \rho_5 \approx 0, \sigma_4 \approx 0$. In the limit of small mixing among the triplets, the model can be treated as a combination of the $Y=0$ (see, e.g.,~\cite{Ross:1975fq,Chankowski:2006hs,Blank:1997qa,Forshaw:2003kh,Chen:2006pb,Chivukula:2007koj,Bandyopadhyay:2020otm,Ashanujjaman:2024lnr}) and the $Y=1$ (see, e.g.,~\cite{Konetschny:1977bn,Schechter:1980gr,Cheng:1980qt,Mohapatra:1980yp,Lazarides:1980nt,Arhrib:2011uy}) models, with the key difference that the triplet VEVs can be sizable, as their contributions to the $\rho$ parameter cancels (partially). 

The particle spectrum contains a doubly charged Higgs $H^{\pm\pm} \equiv \chi^{\pm\pm}$, which originates purely from the $Y=1$ triplet. For small mixing, the remaining mass eigenstates can be identified with the interaction ones as
\begin{align}
 h\approx \phi_r,\; H_{\chi(\zeta)}^0\approx \chi_r(\zeta_r),\; A^0\approx \chi_i,\; H^\pm_{\chi(\zeta)}\approx\chi(\zeta)^\pm,
\end{align}
while $\phi^\pm$($\phi_i$) is approximately the charged (neutral) would-be Goldstone boson. Concerning their masses, the splitting between $m_{H^0_\zeta}$ and $m_{H^\pm_\zeta}$ is of the order of $v_\zeta^2/v$ (i.e.~$m_{H^0_\zeta}\approx m_{H^\pm_\zeta}$), while the splitting among $A^0/H_\chi^0$, $H_\chi^\pm$ and $H^{\pm\pm}$ is of order $v$. To be more specific, with modulus $v_{\chi,\zeta}$ corrections, we have
\begin{align}
& m_{H^{\pm\pm}}^2 - m_{H^\pm_\chi}^2 \approx m_{H^\pm_\chi}^2 - m_{A^0,H^0_\chi}^2.
\end{align}

Via their VEVs, the new Higgses not only acquire di-boson couplings but also induce fermionic couplings for the singly charged and neutral scalars through mixing with the SM Goldstone bosons. The doubly charged Higgs couples to $W^\pm W^\pm$ and $H^\pm_\chi W^\pm$. In the limit of small mixing,
\begin{align}
&H^{\mp\mp}W^\pm_\mu W^\pm_\nu: i 2g^2 v_\chi g_{\mu\nu} \nonumber 
\\
&H^{\mp\mp}H_\chi^\pm W^\pm_\mu: \mp i g (-p^{H^\pm_\chi}_{\mu} + p^{H^{\mp\mp}}_{\mu})\,.
\end{align}
Similarly, the singly charged Higgs $H^\pm_\zeta$ ($H^\pm_\chi$) couples to $WZ$ and $H^0_\zeta W$ ($H^0_\chi W$, $A^0 W$ and $H^{\pm\pm} W^\mp$), and their coupling with $hW$ becomes relevant for sizable mixing between the neutral Higgses $H^0_{\chi,\zeta}$ and $h$, denoted as $\alpha_{\phi\chi}$ and $\alpha_{\phi\zeta}$. They also couple to fermions with a strength of $\approx 2v_{\chi,\zeta}/v Y^f$. Likewise, $H^0_{\chi,\zeta}$ couples to fermions, with a strength proportional to $\alpha_{\phi\chi,\phi\zeta}$. Without mixing, $H^0_\chi$ couples to both $WW$ and $ZZ$ while $H^0_\zeta$ couples only to $WW$. These couplings, normalized to their SM counterparts, are $2\sqrt{2}v_\chi/v$, $4\sqrt{2}v_\chi/v$ and $4v_\zeta/v$, respectively. Also $H^0_{\chi,\zeta}$ has a coupling to $hh$ proportional to $\mu_{\chi\phi}+(\sigma_1+\sigma_2)v_\chi$ [$\mu_{\phi\zeta}-2\sqrt{2}\sigma_3 v_\zeta$] which, via the minimization conditions, is of the order of $v_{\chi,\zeta}$. Finally, the CP-odd Higgs $A^0$ couples to fermions with a strength of $\approx 2\sqrt{2}v_\chi/v Y^f$, and to $hZ$ with a strength of $\approx g/c_W \sqrt{2}v_\chi/v$. Note that these couplings receive corrections due to mixing with the SM Higgs.

\section{Phenomenology}
\label{sec:pheno}

\begin{figure}[t!]
\centering
\includegraphics[scale=0.5]{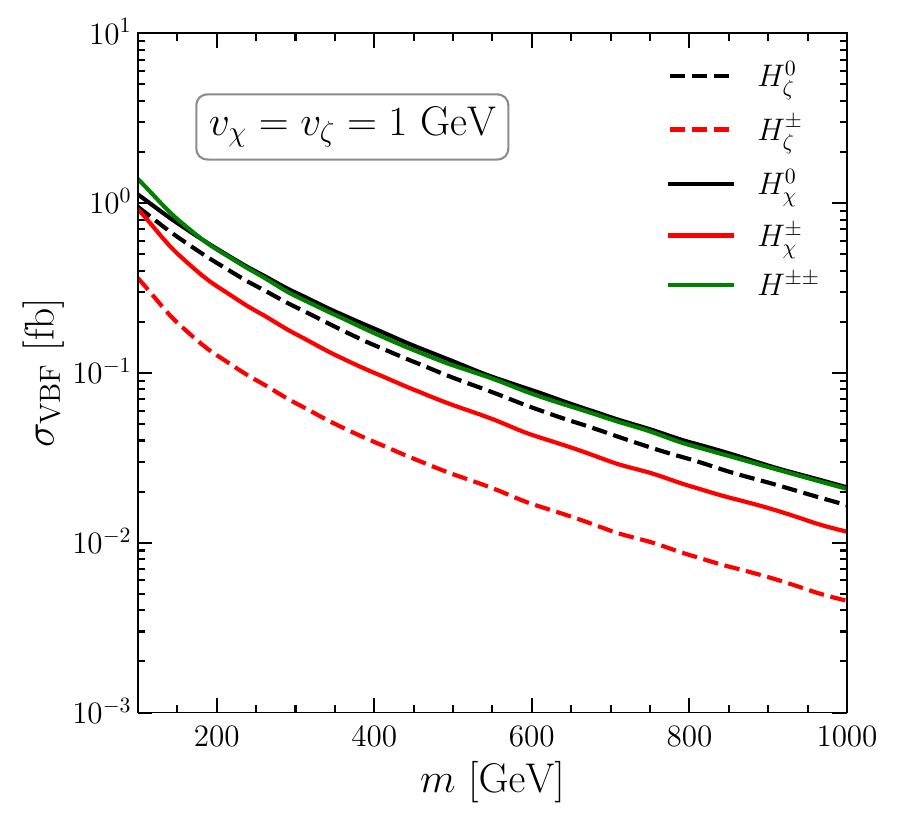}
\caption{VBF production cross sections for the new Higgs states for $v_\chi = v_\zeta = 1$\,GeV at the 13\,TeV LHC, including the NNLO-QCD corrections of Ref.~\cite{LHCHiggsCrossSectionWorkingGroup:2013rie}.}
\label{fig:VBFxs}
\end{figure}

Because we assume that the $Y=0$ and $Y=1$ triplets are decoupled from each other and have only a small mixing with the SM Higgs, to a good approximation, the phenomenology can be studied by considering the effects of the $Y=1$ and $Y=0$ triplet models separately. However, $v_\chi$ and $v_\zeta$ are allowed to be sizable, as their effect on the $W$ mass can cancel.\footnote{The mass splitting among the $\chi$-like Higgses significantly impacts the $W$ mass prediction~\cite{Chun:2012jw}. In contrast, $Y=0$ triplet contribution is suppressed due to ${\cal O}$ (GeV) mass splitting, required by perturbative unitarity and vacuum stability.} Importantly, VBF production of the new Higgs bosons is driven by their di-boson couplings, which are proportional to the VEVs. The production cross sections, in the limit of vanishing mixing, are shown in Fig.~\ref{fig:VBFxs} as a function of their mass for $v_\chi=v_\zeta=1$\,GeV and scale as $v_{\chi}^2$ and $v_{\zeta}^2$ for the $Y=1$ and $Y=0$ triplets, respectively.

\begin{figure*}[t!]
\centering
\includegraphics[scale=0.45]{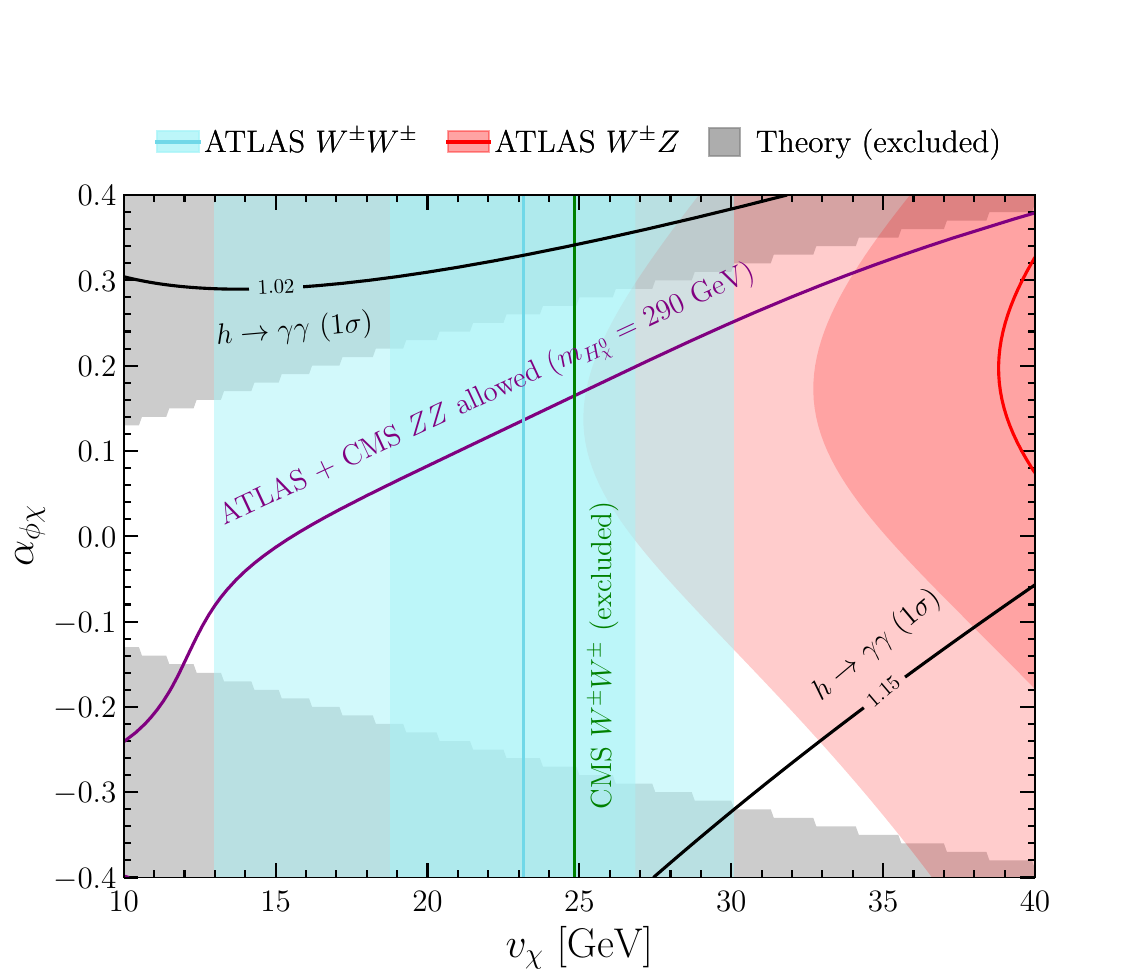}
\includegraphics[scale=0.45]{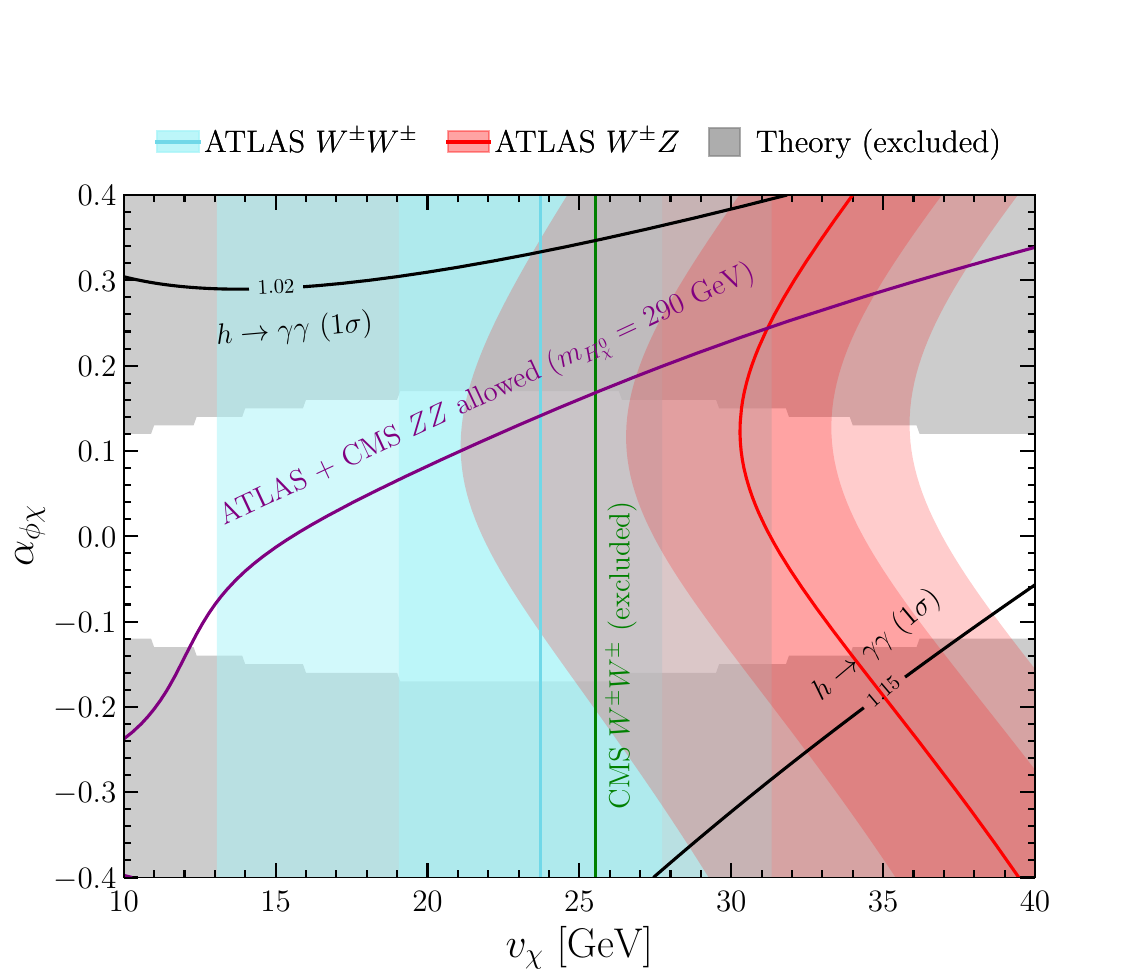}
\caption{Preferred ($1\sigma$ and $2\sigma$) and excluded (95\% CL) regions in the $v_\chi$-$\alpha_{\phi\chi}$ plane. The ATLAS measurement of $W^\pm W^\pm$ prefers the cyan region, while the region above the purple line is allowed by $ZZ$ searches in VBF. Note that both limits are, to a good approximation, independent of the parameters of the $Y=0$ triplet. However, the region (red) preferred by the ATLAS $WZ$ search can be affected by the $Y=0$ triplet parameters in the case of sizable mixing with the $Y=1$ triplet. Also, the region (gray) excluded by theory constraints (vacuum stability and perturbativity) depends on the parameters of this sector. This can be seen by comparing the left plot to the right one. In the left plot, we fix $m_{H^0_\zeta}=$152\,GeV to explain the excess in associated di-photon production, while in the right plot, the charged Higgs states are close in mass (parametrized by $\kappa=0.95$) and have sizable mixing among them, which maximize the effect in $WZ$ (see text for details). The $h\to\gamma\gamma$ contour labels refer to the corresponding signal strength normalized to the SM.}
\label{fig:alpha_vchi}
\end{figure*}

The $W^\pm W^\pm$ excess can only be explained by the doubly-charged Higgs of the $Y=1$ triplet. Accordingly, we can fix $m_{H^{\pm\pm}}\approx 450$\,GeV, for which $\sigma_{\rm VBF}(H^{\pm\pm}) \times {\rm Br}(H^{\pm\pm} \to W^\pm W^\pm) = (72 \pm 25)$~fb is preferred~\cite{ATLAS:2023sua}. Note that this cross section is compatible with the weaker-than-expected 95\% CL limit of $80$~fb from CMS~\cite{CMS:2021wlt}. The best-fit point corresponds to $v_\chi\approx 23$\,GeV (see the cyan band in Fig.~\ref{fig:alpha_vchi}) as long as the decay $H^{\pm\pm}\to H^\pm W^\pm$ is negligible, i.e.,~for Br$(H^{\pm\pm}\to W^\pm W^\pm)\approx 100\%$, which holds for $\Delta m=m_{H^{\pm\pm}}-m_{H^{\pm}}\lessapprox80$\,GeV.

In fact, the $WZ$ excess suggests $m_{H^\pm_\chi}\approx 375$\,GeV, thereby ensuring Br$(H^{\pm\pm}\to W^\pm W^\pm)\approx 100\%$. The best-fit point is $\sigma_{\rm VBF}(H_\chi^\pm)\times {\rm Br}(H_\chi^\pm\to W^\pm Z)\approx 60\pm 20$~fb.\footnote{While ATLAS provides the preferred cross-section for the $W^\pm W^\pm$ excess, they do not provide one for $WZ$. Therefore, we estimated the latter naively from the provided observed and expected limits at 375 GeV.} This is smaller than the 95\% CL upper limit obtained by CMS (which is also weaker than the expected limit)~\cite{CMS:2021wlt}. The dominant decays of $H_\chi^\pm$ are $WZ$, $t\bar b$ and $Wh$. Since the latter depends on the mixing angle of $H^0_\chi$ with the SM Higgs, the red region explaining the $WZ$ excess in Fig.~\ref{fig:alpha_vchi} exhibits its maximal elongation towards small values of $v_\chi$ at $\alpha_{\phi\chi} \approx 0.15$ as this minimizes the $H^\pm Wh$ coupling.

Given the masses of $H^{\pm\pm}$ and $H^\pm_\chi$, the mass of $H^0_\chi$ is approximately determined, i.e.,~$\approx 280+ O( v_{\chi,\zeta})$\,GeV. In addition, $v_\chi$ fixes its production cross section, and its dominant decays ($WW,ZZ$ and $hh$) depend, to a good approximation, only on $\alpha_{\phi\chi}$. The experimental bounds on $\sigma_{\text{VBF}}(H^0_\chi\to ZZ)$ vary in this region of parameter space between 10\,fb and 40\,fb, depending on its exact mass~\cite{ATLAS:2020tlo,CMS:2024vps}. For a mass of 290\,GeV, the ATLAS and CMS limits coincide at $\approx 49$\,fb, and the resulting limit is shown in purple in Fig.~\ref{fig:alpha_vchi} (the region above the purple lines is allowed). Note that this bound suggests a preference for small but positive values of $\alpha_{\phi\chi}$. Further, the constraint from $H^0_\chi\to hh\to 4b$ in VBF~\cite{ATLAS:2020jgy}, which is approximately 1\,pb, is easily satisfied. Finally, the CP-odd Higgs $A^0$---with a mass similar to that of $H^0_\chi$---does not couple to $WW,ZZ$ and is dominantly produced via gluon fusion with a cross section of $\approx 1.5\alpha_{\phi\chi}^2$ times that of a hypothetical SM Higgs with the same mass; it decays mainly to $Zh$ and $b\bar b$. Therefore, the experimental limits are not constraining; even if the branching ratio to $Zh$ were 100\%, the bounds from the corresponding search~\cite{ATLAS:2022enb} would not be violated, given the small production cross section.

\begin{figure*}[t!]
\centering
\includegraphics[scale=0.45]{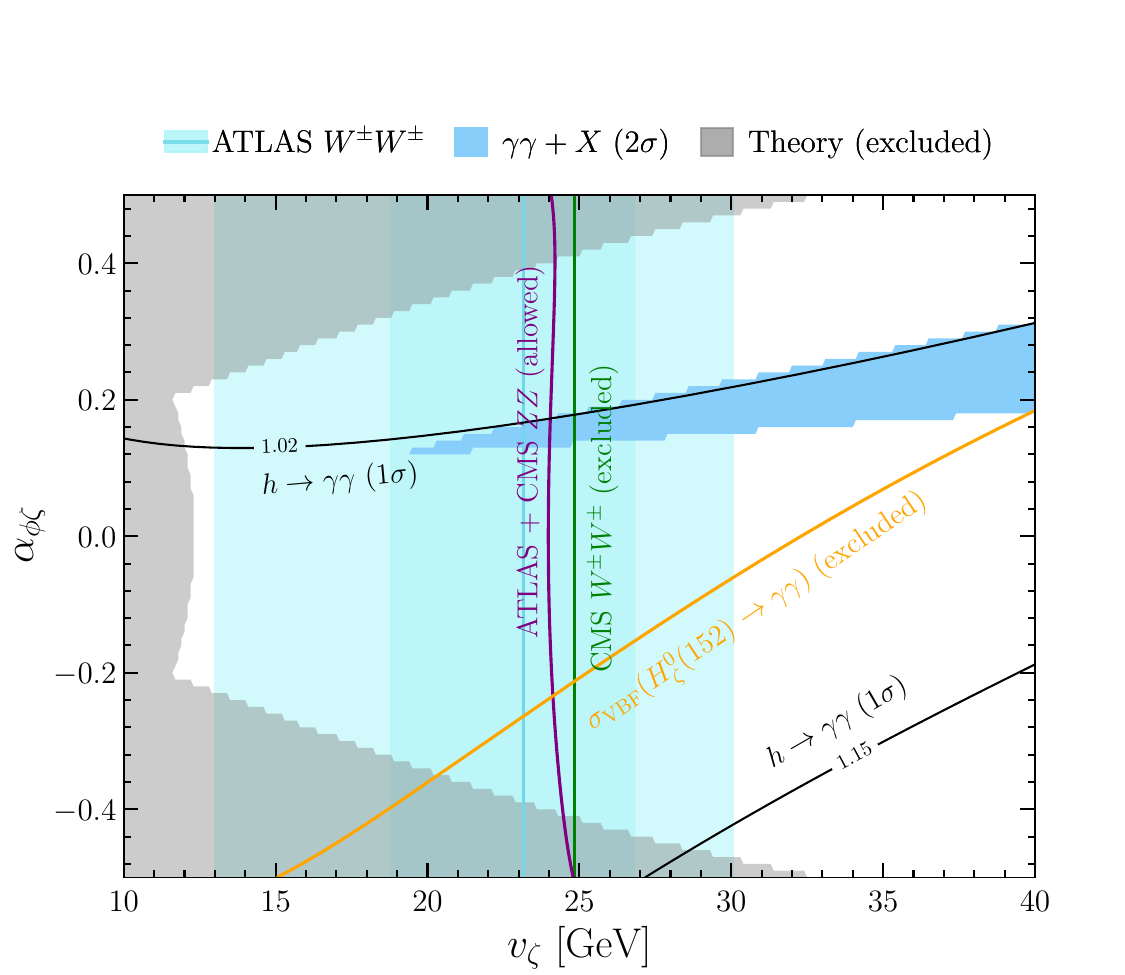}
\includegraphics[scale=0.45]{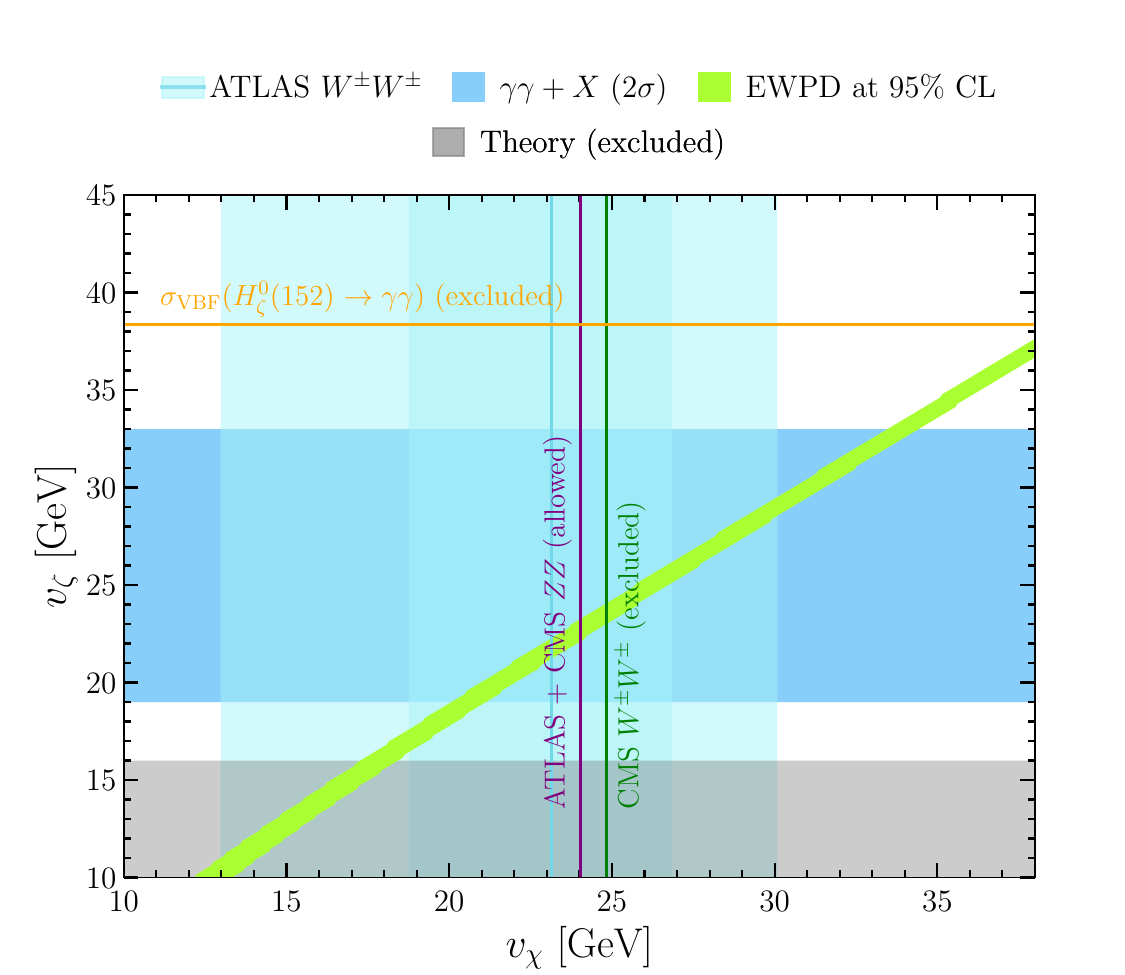}
\caption{Preferred ($1\sigma$ and $2\sigma$) and excluded (95\% CL) regions for $m_{H^0_\zeta}=152\,$GeV in the $v_\zeta$-$\alpha_{\phi\zeta}$ and $v_\chi$-$v_\zeta$ planes for $\alpha_{\phi\chi}=0.15$ (left) and $\alpha_{\phi\zeta}=0.15$ (right), respectively. In the left plot, we assume $v_\chi\approx v_\zeta$, as justified by the green band from EW precision observables in the right plot. The $h\to\gamma\gamma$ contour labels refer to the corresponding signal strenth normalied to the SM.}
\label{fig:152plots}
\end{figure*}

We now discuss the impact of the $Y=0$ triplet. If its quasi-mass-degenerate states $H^0_\zeta$ and $H^\pm_\zeta$ are much heavier or much lighter than $m_{H^\pm_\chi}$, the mixing is small, and the interplay with the bounds discussed so far is limited. In fact, the primary effect is on vacuum stability, which is sensitive to the full scalar potential. The resulting excluded regions are shown in gray in Fig.~\ref{fig:alpha_vchi}.

Only $Y=0$ triplet masses in the range200\,GeV are excluded by multiletpton production at the LHC~\cite{Butterworth:2023rnw}. Therefore, the $Y=0$ triplet can explain the 152\,GeV excess in associated di-photon production~\cite{Ashanujjaman:2024pky,Crivellin:2024uhc,Ashanujjaman:2024lnr}. Because the Drell-Yan production cross section $pp\to W^\pm\to H^0_\zeta H^\pm_\zeta$ and the dominant branching ratios of $H^\pm_\zeta$ are fixed, the excess prefers Br$(H^0_\zeta \to \gamma\gamma)=(0.7\pm 0.18)\%$~\cite{Ashanujjaman:2024lnr}. In Fig.~\ref{fig:152plots}, we show the region in parameter space that gives rise to this branching ratio (and thus explain the excess) for $\alpha_{\phi\chi}=0.15$. Depending on $v_{\zeta}$, $H^0_\zeta$ is also produced in VBF processes, and the resulting di-photon signatures have been studied for the SM Higgs~\cite{ATLAS:2022tnm,CMS:2021kom}. From the sidebands, one can estimate that the cross section of a new Higgs with a mass of 152\,GeV should not exceed 20\% of that of the SM Higgs, excluding the region to the bottom-right of the orange line in Fig.~\ref{fig:152plots} (left).

We next project this scenario onto the $v_\chi$-$v_\zeta$ plane in the right plot in Fig.~\ref{fig:152plots}. Here, the green band illustrates the degree of cancellation in the $W$ mass between the tree-level contribution due to $v_\chi$, $v_\zeta$ and the loop contribution induced by the mass splitting among the $Y=1$ components needed to respect the EW precision bounds. Note that, contrary to the pure $Y=0$ triplet model, where there are tensions between the preferred branching ratio from the di-photon excess, vacuum stability, and SM Higgs signal strength measurements, the gGMM can resolve these issues.

Finally, if one allows for similar masses of the two charged Higgses, one has sizable mixing among them due to $v_{\chi}$ and $v_{\zeta}$. This mixing enhances Br$(H^\pm_\chi \to WZ)$, thereby enlarging the overlap of the corresponding preferred region with that of $W^\pm W^\pm$. For this, we consider $m_{H^{\pm\pm}} = 450$\,GeV, $m_{H^\pm_\chi} = 375$\,GeV, $m_{H^\pm_\zeta} = \kappa \times \frac{v_\phi}{\sqrt{v_\phi^2 + 8 v_\chi^2}} m_{H^\pm_\chi}$\,GeV; $\kappa \to 1$ leads to maximal mixing. The resulting preferred and excluded regions for $\kappa = 0.95$ are shown in the right plot of Fig.~\ref{fig:alpha_vchi}.

\section{Conclusions and Outlook}
\label{sec:conclusion}

In this article, we explored the gGMM in the context of the indications for new scalars in VBF production in $W^\pm W^\pm$ and $WZ$ at $\approx450$\,GeV and $\approx375$\,GeV with local significance of $3.3\sigma$ and $2.8\sigma$, respectively. Working in the limit of small mixing angles, we fixed the masses of the $Y=1$ triplet states to these values, which predicts its neutral components to be around $280$\,GeV. We showed that the resulting constraints from $ZZ$ in VBF, SM Higgs signal strength, as well as the theoretical bounds from vacuum stability and perturbative unitarity, can be respected for VEVs capable of providing sufficiently high production cross sections (see Fig.~\ref{fig:alpha_vchi}). However, the regions preferred by the $W^\pm W^\pm$ and $WZ$ channels only overlap at the $2\sigma$ level.

Furthermore, we showed that the neutral component of the $Y=0$ triplet can account for the di-photon excess in associated production channels without violating the previously mentioned bounds or the $\gamma\gamma$ signal in VBF or EW precision data (see Fig.~\ref{fig:152plots}). In fact, the agreement with data is better than in the $\Delta$SM (SM extended only with the $Y=0$ triplet), as higher values of Br$(H^0_\zeta\to\gamma\gamma)$ can be obtained without conflicting with vacuum stability or the SM Higgs signal strength measurements. Alternatively, if $H^\pm_\zeta$ has a mass close to $375$\,GeV, the $WZ$ cross section can be enhanced such that a better agreement with the $W^{\pm}W^\pm$ signal in VBF is obtained.

\begin{acknowledgments}
This research was supported by the Deutsche Forschungsgemeinschaft (DFG, German Research Foundation) under grant 396021762 - TRR 257. The work of A.C.~is supported by a professorship grant from the Swiss National Science Foundation (No.\ PP00P21\_76884). S.P.M. would like to acknowledge the support of the research office of the University of the Witwatersrand.
\end{acknowledgments}

\bibliographystyle{utphys}
\bibliography{PRL}

\providecommand{\href}[2]{#2}\begingroup\raggedright\begin{thebibliography}{10}

\bibitem{ParticleDataGroup:2022pth}
{\bfseries Particle Data Group} Collaboration, R.~L. Workman {\em et~al.},
  ``{Review of Particle Physics},''
  \href{http://dx.doi.org/10.1093/ptep/ptac097}{{\em PTEP} {\bfseries 2022}
  (2022) 083C01}.

\bibitem{Aad:2012tfa}
{\bfseries ATLAS} Collaboration, G.~Aad {\em et~al.}, ``{Observation of a new
  particle in the search for the Standard Model Higgs boson with the ATLAS
  detector at the LHC},''
  \href{http://dx.doi.org/10.1016/j.physletb.2012.08.020}{{\em Phys. Lett. B}
  {\bfseries 716} (2012) 1--29},
  \href{http://arxiv.org/abs/1207.7214}{{\ttfamily arXiv:1207.7214 [hep-ex]}}.

\bibitem{Chatrchyan:2012ufa}
{\bfseries CMS} Collaboration, S.~Chatrchyan {\em et~al.}, ``{Observation of a
  New Boson at a Mass of 125 GeV with the CMS Experiment at the LHC},''
  \href{http://dx.doi.org/10.1016/j.physletb.2012.08.021}{{\em Phys. Lett. B}
  {\bfseries 716} (2012) 30--61},
  \href{http://arxiv.org/abs/1207.7235}{{\ttfamily arXiv:1207.7235 [hep-ex]}}.

\bibitem{ATLAS:2022vkf}
{\bfseries ATLAS} Collaboration, G.~Aad {\em et~al.}, ``{A detailed map of
  Higgs boson interactions by the ATLAS experiment ten years after the
  discovery},'' \href{http://dx.doi.org/10.1038/s41586-022-04893-w}{{\em
  Nature} {\bfseries 607} no.~7917, (2022) 52--59},
  \href{http://arxiv.org/abs/2207.00092}{{\ttfamily arXiv:2207.00092
  [hep-ex]}}. [Erratum: Nature 612, E24 (2022)].

\bibitem{CMS:2022dwd}
{\bfseries CMS} Collaboration, A.~Tumasyan {\em et~al.}, ``{A portrait of the
  Higgs boson by the CMS experiment ten years after the discovery.},''
  \href{http://dx.doi.org/10.1038/s41586-022-04892-x}{{\em Nature} {\bfseries
  607} no.~7917, (2022) 60--68},
  \href{http://arxiv.org/abs/2207.00043}{{\ttfamily arXiv:2207.00043
  [hep-ex]}}. [Erratum: Nature 623, (2023)].

\bibitem{Haller:2022eyb}
J.~Haller, A.~Hoecker, R.~Kogler, K.~M\"onig, and J.~Stelzer, ``{Status of the
  global electroweak fit with Gfitter in the light of new precision
  measurements},'' \href{http://dx.doi.org/10.22323/1.414.0897}{{\em PoS}
  {\bfseries ICHEP2022} (11, 2022) 897},
  \href{http://arxiv.org/abs/2211.07665}{{\ttfamily arXiv:2211.07665
  [hep-ph]}}.

\bibitem{deBlas:2021wap}
J.~de~Blas, M.~Ciuchini, E.~Franco, A.~Goncalves, S.~Mishima, M.~Pierini,
  L.~Reina, and L.~Silvestrini, ``{Global analysis of electroweak data in the
  Standard Model},'' \href{http://dx.doi.org/10.1103/PhysRevD.106.033003}{{\em
  Phys. Rev. D} {\bfseries 106} no.~3, (2022) 033003},
  \href{http://arxiv.org/abs/2112.07274}{{\ttfamily arXiv:2112.07274
  [hep-ph]}}.

\bibitem{ATLAS:2022zuc}
{\bfseries ATLAS} Collaboration, G.~Aad {\em et~al.}, ``{Search for resonant WZ
  production in the fully leptonic final state in proton\textendash{}proton
  collisions at $\mathbf {\sqrt{s} = 13}$~TeV with the ATLAS detector},''
  \href{http://dx.doi.org/10.1140/epjc/s10052-023-11437-7}{{\em Eur. Phys. J.
  C} {\bfseries 83} no.~7, (2023) 633},
  \href{http://arxiv.org/abs/2207.03925}{{\ttfamily arXiv:2207.03925
  [hep-ex]}}.

\bibitem{ATLAS:2023sua}
{\bfseries ATLAS} Collaboration, G.~Aad {\em et~al.}, ``{Measurement and
  interpretation of same-sign W boson pair production in association with two
  jets in pp collisions at $ \sqrt{s} $ = 13 TeV with the ATLAS detector},''
  \href{http://dx.doi.org/10.1007/JHEP04(2024)026}{{\em JHEP} {\bfseries 04}
  (2024) 026}, \href{http://arxiv.org/abs/2312.00420}{{\ttfamily
  arXiv:2312.00420 [hep-ex]}}.

\bibitem{CMS:2021wlt}
{\bfseries CMS} Collaboration, A.~M. Sirunyan {\em et~al.}, ``{Search for
  charged Higgs bosons produced in vector boson fusion processes and decaying
  into vector boson pairs in proton\textendash{}proton collisions at $\sqrt{s}
  = 13\,{\text {TeV}} $},''
  \href{http://dx.doi.org/10.1140/epjc/s10052-021-09472-3}{{\em Eur. Phys. J.
  C} {\bfseries 81} no.~8, (2021) 723},
  \href{http://arxiv.org/abs/2104.04762}{{\ttfamily arXiv:2104.04762
  [hep-ex]}}.

\bibitem{Georgi:1985nv}
H.~Georgi and M.~Machacek, ``{DOUBLY CHARGED HIGGS BOSONS},''
  \href{http://dx.doi.org/10.1016/0550-3213(85)90325-6}{{\em Nucl. Phys. B}
  {\bfseries 262} (1985) 463--477}.

\bibitem{Chanowitz:1985ug}
M.~S. Chanowitz and M.~Golden, ``{Higgs Boson Triplets With M($W$) = M($Z$)
  $\cos \theta _\omega$},''
  \href{http://dx.doi.org/10.1016/0370-2693(85)90700-2}{{\em Phys. Lett. B}
  {\bfseries 165} (1985) 105--108}.

\bibitem{Gunion:1989ci}
J.~F. Gunion, R.~Vega, and J.~Wudka, ``{Higgs triplets in the standard
  model},'' \href{http://dx.doi.org/10.1103/PhysRevD.42.1673}{{\em Phys. Rev.
  D} {\bfseries 42} (1990) 1673--1691}.

\bibitem{Godfrey:2010qb}
S.~Godfrey and K.~Moats, ``{Exploring Higgs Triplet Models via Vector Boson
  Scattering at the LHC},''
  \href{http://dx.doi.org/10.1103/PhysRevD.81.075026}{{\em Phys. Rev. D}
  {\bfseries 81} (2010) 075026},
  \href{http://arxiv.org/abs/1003.3033}{{\ttfamily arXiv:1003.3033 [hep-ph]}}.

\bibitem{Chiang:2012cn}
C.-W. Chiang and K.~Yagyu, ``{Testing the custodial symmetry in the Higgs
  sector of the Georgi-Machacek model},''
  \href{http://dx.doi.org/10.1007/JHEP01(2013)026}{{\em JHEP} {\bfseries 01}
  (2013) 026}, \href{http://arxiv.org/abs/1211.2658}{{\ttfamily arXiv:1211.2658
  [hep-ph]}}.

\bibitem{Englert:2013zpa}
C.~Englert, E.~Re, and M.~Spannowsky, ``{Triplet Higgs boson collider
  phenomenology after the LHC},''
  \href{http://dx.doi.org/10.1103/PhysRevD.87.095014}{{\em Phys. Rev. D}
  {\bfseries 87} no.~9, (2013) 095014},
  \href{http://arxiv.org/abs/1302.6505}{{\ttfamily arXiv:1302.6505 [hep-ph]}}.

\bibitem{Hartling:2014zca}
K.~Hartling, K.~Kumar, and H.~E. Logan, ``{The decoupling limit in the
  Georgi-Machacek model},''
  \href{http://dx.doi.org/10.1103/PhysRevD.90.015007}{{\em Phys. Rev. D}
  {\bfseries 90} no.~1, (2014) 015007},
  \href{http://arxiv.org/abs/1404.2640}{{\ttfamily arXiv:1404.2640 [hep-ph]}}.

\bibitem{Godunov:2014waa}
S.~I. Godunov, M.~I. Vysotsky, and E.~V. Zhemchugov, ``{Double Higgs production
  at LHC, see-saw type II and Georgi-Machacek model},''
  \href{http://dx.doi.org/10.1134/S1063776115030073}{{\em J. Exp. Theor. Phys.}
  {\bfseries 120} no.~3, (2015) 369--375},
  \href{http://arxiv.org/abs/1408.0184}{{\ttfamily arXiv:1408.0184 [hep-ph]}}.

\bibitem{Logan:2015xpa}
H.~E. Logan and V.~Rentala, ``{All the generalized Georgi-Machacek models},''
  \href{http://dx.doi.org/10.1103/PhysRevD.92.075011}{{\em Phys. Rev. D}
  {\bfseries 92} no.~7, (2015) 075011},
  \href{http://arxiv.org/abs/1502.01275}{{\ttfamily arXiv:1502.01275
  [hep-ph]}}.

\bibitem{Chiang:2015kka}
C.-W. Chiang and K.~Tsumura, ``{Properties and searches of the exotic neutral
  Higgs bosons in the Georgi-Machacek model},''
  \href{http://dx.doi.org/10.1007/JHEP04(2015)113}{{\em JHEP} {\bfseries 04}
  (2015) 113}, \href{http://arxiv.org/abs/1501.04257}{{\ttfamily
  arXiv:1501.04257 [hep-ph]}}.

\bibitem{Chiang:2015amq}
C.-W. Chiang, A.-L. Kuo, and T.~Yamada, ``{Searches of exotic Higgs bosons in
  general mass spectra of the Georgi-Machacek model at the LHC},''
  \href{http://dx.doi.org/10.1007/JHEP01(2016)120}{{\em JHEP} {\bfseries 01}
  (2016) 120}, \href{http://arxiv.org/abs/1511.00865}{{\ttfamily
  arXiv:1511.00865 [hep-ph]}}.

\bibitem{Chen:2023bqr}
T.-K. Chen, C.-W. Chiang, S.~Heinemeyer, and G.~Weiglein, ``{95~GeV Higgs boson
  in the Georgi-Machacek model},''
  \href{http://dx.doi.org/10.1103/PhysRevD.109.075043}{{\em Phys. Rev. D}
  {\bfseries 109} no.~7, (2024) 075043},
  \href{http://arxiv.org/abs/2312.13239}{{\ttfamily arXiv:2312.13239
  [hep-ph]}}.

\bibitem{Ghosh:2023izq}
S.~Ghosh, ``{Charged Higgs Decay to $W^{\pm}$ and Heavy Neutral Higgs Decaying
  into $\tau^+ \tau^-$ in Georgi-Machacek Model at LHC},''
  \href{http://dx.doi.org/10.31526/lhep.2024.518}{{\em LHEP} {\bfseries 2024}
  (2024) 518}, \href{http://arxiv.org/abs/2311.15405}{{\ttfamily
  arXiv:2311.15405 [hep-ph]}}.

\bibitem{Chakraborti:2023mya}
M.~Chakraborti, D.~Das, N.~Ghosh, S.~Mukherjee, and I.~Saha, ``{New physics
  implications of vector boson fusion searches exemplified through the
  Georgi-Machacek model},''
  \href{http://dx.doi.org/10.1103/PhysRevD.109.015016}{{\em Phys. Rev. D}
  {\bfseries 109} no.~1, (2024) 015016},
  \href{http://arxiv.org/abs/2308.02384}{{\ttfamily arXiv:2308.02384
  [hep-ph]}}.

\bibitem{Ahriche:2022aoj}
A.~Ahriche, ``{Constraining the Georgi-Machacek model with a light Higgs
  boson},'' \href{http://dx.doi.org/10.1103/PhysRevD.107.015006}{{\em Phys.
  Rev. D} {\bfseries 107} no.~1, (2023) 015006},
  \href{http://arxiv.org/abs/2212.11579}{{\ttfamily arXiv:2212.11579
  [hep-ph]}}.

\bibitem{Ghosh:2022bvz}
R.~Ghosh and B.~Mukhopadhyaya, ``{Some new observations for the Georgi-Machacek
  scenario with triplet Higgs scalars},''
  \href{http://dx.doi.org/10.1103/PhysRevD.107.035031}{{\em Phys. Rev. D}
  {\bfseries 107} no.~3, (2023) 035031},
  \href{http://arxiv.org/abs/2212.11688}{{\ttfamily arXiv:2212.11688
  [hep-ph]}}.

\bibitem{deLima:2022yvn}
C.~H. de~Lima and H.~E. Logan, ``{Unavoidable Higgs coupling deviations in the
  $Z_2$-symmetric Georgi-Machacek model},''
  \href{http://dx.doi.org/10.1103/PhysRevD.106.115020}{{\em Phys. Rev. D}
  {\bfseries 106} no.~11, (2022) 115020},
  \href{http://arxiv.org/abs/2209.08393}{{\ttfamily arXiv:2209.08393
  [hep-ph]}}.

\bibitem{Chen:2022zsh}
T.-K. Chen, C.-W. Chiang, C.-T. Huang, and B.-Q. Lu, ``{Updated constraints on
  the Georgi-Machacek model and its electroweak phase transition and associated
  gravitational waves},''
  \href{http://dx.doi.org/10.1103/PhysRevD.106.055019}{{\em Phys. Rev. D}
  {\bfseries 106} no.~5, (2022) 055019},
  \href{http://arxiv.org/abs/2205.02064}{{\ttfamily arXiv:2205.02064
  [hep-ph]}}.

\bibitem{Chen:2022ocr}
T.-K. Chen, C.-W. Chiang, and K.~Yagyu, ``{Explanation of the $W$ mass shift at
  CDF II in the extended Georgi-Machacek model},''
  \href{http://dx.doi.org/10.1103/PhysRevD.106.055035}{{\em Phys. Rev. D}
  {\bfseries 106} no.~5, (2022) 055035},
  \href{http://arxiv.org/abs/2204.12898}{{\ttfamily arXiv:2204.12898
  [hep-ph]}}.

\bibitem{Wang:2022okq}
C.~Wang, J.-Q. Tao, M.~A. Shahzad, G.-M. Chen, and S.~Gascon-Shotkin, ``{Search
  for a lighter neutral custodial fiveplet scalar in the Georgi-Machacek model
  *},'' \href{http://dx.doi.org/10.1088/1674-1137/ac6cd3}{{\em Chin. Phys. C}
  {\bfseries 46} no.~8, (2022) 083107},
  \href{http://arxiv.org/abs/2204.09198}{{\ttfamily arXiv:2204.09198
  [hep-ph]}}.

\bibitem{Zhu:2020bew}
J.-W. Zhu, R.-Y. Zhang, W.-G. Ma, Q.~Yang, M.-M. Long, and Y.~Jiang, ``{Search
  for SU(2)$_V$ singlet Higgs boson in the Georgi\textendash{}Machacek model at
  the LHC},'' \href{http://dx.doi.org/10.1088/1361-6471/abaddf}{{\em J. Phys.
  G} {\bfseries 47} no.~12, (2020) 125005}.

\bibitem{Ismail:2020kqz}
A.~Ismail, B.~Keeshan, H.~E. Logan, and Y.~Wu, ``{Benchmark for LHC searches
  for low-mass custodial fiveplet scalars in the Georgi-Machacek model},''
  \href{http://dx.doi.org/10.1103/PhysRevD.103.095010}{{\em Phys. Rev. D}
  {\bfseries 103} no.~9, (2021) 095010},
  \href{http://arxiv.org/abs/2003.05536}{{\ttfamily arXiv:2003.05536
  [hep-ph]}}.

\bibitem{Das:2018vkv}
D.~Das and I.~Saha, ``{Cornering variants of the Georgi-Machacek model using
  Higgs precision data},''
  \href{http://dx.doi.org/10.1103/PhysRevD.98.095010}{{\em Phys. Rev. D}
  {\bfseries 98} no.~9, (2018) 095010},
  \href{http://arxiv.org/abs/1811.00979}{{\ttfamily arXiv:1811.00979
  [hep-ph]}}.

\bibitem{Yang:2018pzt}
Q.~Yang, R.-Y. Zhang, W.-G. Ma, Y.~Jiang, X.-Z. Li, and H.~Sun,
  ``{$H_5^{\pm\pm}h_0$ production via vector-boson fusion in the
  Georgi-Machacek model at hadron colliders},''
  \href{http://dx.doi.org/10.1103/PhysRevD.98.055034}{{\em Phys. Rev. D}
  {\bfseries 98} no.~5, (2018) 055034},
  \href{http://arxiv.org/abs/1810.08965}{{\ttfamily arXiv:1810.08965
  [hep-ph]}}.

\bibitem{Keeshan:2018ypw}
B.~Keeshan, H.~E. Logan, and T.~Pilkington, ``{Custodial symmetry violation in
  the Georgi-Machacek model},''
  \href{http://dx.doi.org/10.1103/PhysRevD.102.015001}{{\em Phys. Rev. D}
  {\bfseries 102} no.~1, (2020) 015001},
  \href{http://arxiv.org/abs/1807.11511}{{\ttfamily arXiv:1807.11511
  [hep-ph]}}.

\bibitem{Chiang:2018cgb}
C.-W. Chiang, G.~Cottin, and O.~Eberhardt, ``{Global fits in the
  Georgi-Machacek model},''
  \href{http://dx.doi.org/10.1103/PhysRevD.99.015001}{{\em Phys. Rev. D}
  {\bfseries 99} no.~1, (2019) 015001},
  \href{http://arxiv.org/abs/1807.10660}{{\ttfamily arXiv:1807.10660
  [hep-ph]}}.

\bibitem{Logan:2017jpr}
H.~E. Logan and M.~B. Reimer, ``{Characterizing a benchmark scenario for heavy
  Higgs boson searches in the Georgi-Machacek model},''
  \href{http://dx.doi.org/10.1103/PhysRevD.96.095029}{{\em Phys. Rev. D}
  {\bfseries 96} no.~9, (2017) 095029},
  \href{http://arxiv.org/abs/1709.01883}{{\ttfamily arXiv:1709.01883
  [hep-ph]}}.

\bibitem{Chang:2017niy}
J.~Chang, C.-R. Chen, and C.-W. Chiang, ``{Higgs boson pair productions in the
  Georgi-Machacek model at the LHC},''
  \href{http://dx.doi.org/10.1007/JHEP03(2017)137}{{\em JHEP} {\bfseries 03}
  (2017) 137}, \href{http://arxiv.org/abs/1701.06291}{{\ttfamily
  arXiv:1701.06291 [hep-ph]}}.

\bibitem{Kanemura:2024ezz}
S.~Kanemura and Y.~Mura, ``{Loop induced H$^{\pm}$W$^{\mp}$Z vertices in the
  general two Higgs doublet model with CP violation},''
  \href{http://dx.doi.org/10.1007/JHEP10(2024)041}{{\em JHEP} {\bfseries 10}
  (2024) 041}, \href{http://arxiv.org/abs/2408.06863}{{\ttfamily
  arXiv:2408.06863 [hep-ph]}}.

\bibitem{LeYaouanc:2023zvi}
A.~Le~Yaouanc and F.~Richard, ``{As a consequence of $H(650) \to W^+ W^-/ZZ$,
  one predicts $H^{++}\to W^+W^+$ and $H^+\to ZW^+$, as indicated by LHC
  data},'' in {\em {2nd ECFA Workshop on $e^+e^-$ Higgs/EW/Top Factories}}.
\newblock 8, 2023.
\newblock \href{http://arxiv.org/abs/2308.12180}{{\ttfamily arXiv:2308.12180
  [hep-ph]}}.

\bibitem{Blasi:2017xmc}
S.~Blasi, S.~De~Curtis, and K.~Yagyu, ``{Effects of custodial symmetry breaking
  in the Georgi-Machacek model at high energies},''
  \href{http://dx.doi.org/10.1103/PhysRevD.96.015001}{{\em Phys. Rev. D}
  {\bfseries 96} no.~1, (2017) 015001},
  \href{http://arxiv.org/abs/1704.08512}{{\ttfamily arXiv:1704.08512
  [hep-ph]}}.

\bibitem{Chen:2023ins}
T.-K. Chen, C.-W. Chiang, and K.~Yagyu, ``{CP violation in a model with Higgs
  triplets},'' \href{http://dx.doi.org/10.1007/JHEP06(2023)069}{{\em JHEP}
  {\bfseries 06} (2023) 069}, \href{http://arxiv.org/abs/2303.09294}{{\ttfamily
  arXiv:2303.09294 [hep-ph]}}. [Erratum: JHEP 07, 169 (2023)].

\bibitem{Crivellin:2021ubm}
A.~Crivellin, Y.~Fang, O.~Fischer, S.~Bhattacharya, M.~Kumar, E.~Malwa,
  B.~Mellado, N.~Rapheeha, X.~Ruan, and Q.~Sha, ``{Accumulating evidence for
  the associated production of a new Higgs boson at the LHC},''
  \href{http://dx.doi.org/10.1103/PhysRevD.108.115031}{{\em Phys. Rev. D}
  {\bfseries 108} no.~11, (2023) 115031},
  \href{http://arxiv.org/abs/2109.02650}{{\ttfamily arXiv:2109.02650
  [hep-ph]}}.

\bibitem{Bhattacharya:2023lmu}
S.~Bhattacharya, G.~Coloretti, A.~Crivellin, S.-E. Dahbi, Y.~Fang, M.~Kumar,
  and B.~Mellado, ``{Growing Excesses of New Scalars at the Electroweak
  Scale},'' \href{http://arxiv.org/abs/2306.17209}{{\ttfamily arXiv:2306.17209
  [hep-ph]}}.

\bibitem{Ashanujjaman:2024pky}
S.~Ashanujjaman, S.~Banik, G.~Coloretti, A.~Crivellin, S.~P. Maharathy, and
  B.~Mellado, ``{Explaining the $\gamma \gamma$ + X excesses at $\approx$151.5
  GeV via the Drell-Yan production of a Higgs triplet},''
  \href{http://dx.doi.org/10.1016/j.physletb.2025.139298}{{\em Phys. Lett. B}
  {\bfseries 862} (2025) 139298},
  \href{http://arxiv.org/abs/2402.00101}{{\ttfamily arXiv:2402.00101
  [hep-ph]}}.

\bibitem{Crivellin:2024uhc}
A.~Crivellin, S.~Ashanujjaman, S.~Banik, G.~Coloretti, S.~P. Maharathy, and
  B.~Mellado, ``{Growing Evidence for a Higgs Triplet},''
  \href{http://arxiv.org/abs/2404.14492}{{\ttfamily arXiv:2404.14492
  [hep-ph]}}.

\bibitem{Banik:2024ugs}
S.~Banik, G.~Coloretti, A.~Crivellin, and H.~E. Haber, ``{Correlating $A\to
  \gamma\gamma$ with EDMs in the 2HDM in light of the diphoton excesses at 95
  GeV and 152 GeV},'' \href{http://arxiv.org/abs/2412.00523}{{\ttfamily
  arXiv:2412.00523 [hep-ph]}}.

\bibitem{Ashanujjaman:2024lnr}
S.~Ashanujjaman, S.~Banik, G.~Coloretti, A.~Crivellin, S.~P. Maharathy, and
  B.~Mellado, ``{Anatomy of the Real Higgs Triplet Model},''
  \href{http://arxiv.org/abs/2411.18618}{{\ttfamily arXiv:2411.18618
  [hep-ph]}}.

\bibitem{ATLAS:2023omk}
{\bfseries ATLAS} Collaboration, G.~Aad {\em et~al.}, ``{Model-independent
  search for the presence of new physics in events including
  $H\rightarrow\gamma\gamma$ with $\sqrt{s}$ = 13 TeV pp data recorded by the
  ATLAS detector at the LHC},''
  \href{http://dx.doi.org/10.1007/JHEP07(2023)176}{{\em JHEP} {\bfseries 07}
  (2023) 176}, \href{http://arxiv.org/abs/2301.10486}{{\ttfamily
  arXiv:2301.10486 [hep-ex]}}.

\bibitem{ATLAS:2024lhu}
{\bfseries ATLAS} Collaboration, G.~Aad {\em et~al.}, ``{Search for
  non-resonant Higgs boson pair production in final states with leptons, taus,
  and photons in pp collisions at $ \sqrt{s} $ = 13 TeV with the ATLAS
  detector},'' \href{http://dx.doi.org/10.1007/JHEP08(2024)164}{{\em JHEP}
  {\bfseries 08} (2024) 164}, \href{http://arxiv.org/abs/2405.20040}{{\ttfamily
  arXiv:2405.20040 [hep-ex]}}.

\bibitem{vonBuddenbrock:2017gvy}
S.~von Buddenbrock, A.~S. Cornell, A.~Fadol, M.~Kumar, B.~Mellado, and X.~Ruan,
  ``{Multi-lepton signatures of additional scalar bosons beyond the Standard
  Model at the LHC},'' \href{http://dx.doi.org/10.1088/1361-6471/aae3d6}{{\em
  J. Phys. G} {\bfseries 45} no.~11, (2018) 115003},
  \href{http://arxiv.org/abs/1711.07874}{{\ttfamily arXiv:1711.07874
  [hep-ph]}}.

\bibitem{Buddenbrock:2019tua}
S.~Buddenbrock, A.~S. Cornell, Y.~Fang, A.~Fadol~Mohammed, M.~Kumar,
  B.~Mellado, and K.~G. Tomiwa, ``{The emergence of multi-lepton anomalies at
  the LHC and their compatibility with new physics at the EW scale},''
  \href{http://dx.doi.org/10.1007/JHEP10(2019)157}{{\em JHEP} {\bfseries 10}
  (2019) 157}, \href{http://arxiv.org/abs/1901.05300}{{\ttfamily
  arXiv:1901.05300 [hep-ph]}}.

\bibitem{Coloretti:2023wng}
G.~Coloretti, A.~Crivellin, S.~Bhattacharya, and B.~Mellado, ``{Searching for
  low-mass resonances decaying into W bosons},''
  \href{http://dx.doi.org/10.1103/PhysRevD.108.035026}{{\em Phys. Rev. D}
  {\bfseries 108} no.~3, (2023) 035026},
  \href{http://arxiv.org/abs/2302.07276}{{\ttfamily arXiv:2302.07276
  [hep-ph]}}.

\bibitem{Banik:2023vxa}
S.~Banik, G.~Coloretti, A.~Crivellin, and B.~Mellado, ``{Uncovering new Higgses
  in the LHC analyses of differential $ t\overline{t} $ cross sections},''
  \href{http://dx.doi.org/10.1007/JHEP01(2025)155}{{\em JHEP} {\bfseries 01}
  (2025) 155}, \href{http://arxiv.org/abs/2308.07953}{{\ttfamily
  arXiv:2308.07953 [hep-ph]}}.

\bibitem{Fischer:2021sqw}
O.~Fischer {\em et~al.}, ``{Unveiling hidden physics at the LHC},''
  \href{http://dx.doi.org/10.1140/epjc/s10052-022-10541-4}{{\em Eur. Phys. J.
  C} {\bfseries 82} no.~8, (2022) 665},
  \href{http://arxiv.org/abs/2109.06065}{{\ttfamily arXiv:2109.06065
  [hep-ph]}}.

\bibitem{Crivellin:2023zui}
A.~Crivellin and B.~Mellado, ``{Anomalies in particle physics and their
  implications for physics beyond the standard model},''
  \href{http://dx.doi.org/10.1038/s42254-024-00703-6}{{\em Nature Rev. Phys.}
  {\bfseries 6} no.~5, (2024) 294--309},
  \href{http://arxiv.org/abs/2309.03870}{{\ttfamily arXiv:2309.03870
  [hep-ph]}}.

\bibitem{Richard:2021ovc}
F.~Richard, ``{A Georgi-Machacek Interpretation of the Associate Production of
  a Neutral Scalar with Mass around 151 GeV},'' in {\em {ILC Workshop on
  Potential Experiments}}.
\newblock 12, 2021.
\newblock \href{http://arxiv.org/abs/2112.07982}{{\ttfamily arXiv:2112.07982
  [hep-ph]}}.

\bibitem{Ross:1975fq}
D.~A. Ross and M.~J.~G. Veltman, ``{Neutral Currents in Neutrino
  Experiments},'' \href{http://dx.doi.org/10.1016/0550-3213(75)90485-X}{{\em
  Nucl. Phys. B} {\bfseries 95} (1975) 135--147}.

\bibitem{Chankowski:2006hs}
P.~H. Chankowski, S.~Pokorski, and J.~Wagner, ``{(Non)decoupling of the Higgs
  triplet effects},''
  \href{http://dx.doi.org/10.1140/epjc/s10052-007-0259-x}{{\em Eur. Phys. J. C}
  {\bfseries 50} (2007) 919--933},
  \href{http://arxiv.org/abs/hep-ph/0605302}{{\ttfamily arXiv:hep-ph/0605302}}.

\bibitem{Blank:1997qa}
T.~Blank and W.~Hollik, ``{Precision observables in $SU(2) \times U(1)$ models
  with an additional Higgs triplet},''
  \href{http://dx.doi.org/10.Ashanujjaman:2024lnr1016/S0550-3213(97)00785-2}{{\em
  Nucl. Phys. B} {\bfseries 514} (1998) 113--134},
  \href{http://arxiv.org/abs/hep-ph/9703392}{{\ttfamily arXiv:hep-ph/9703392}}.

\bibitem{Forshaw:2003kh}
J.~R. Forshaw, A.~Sabio~Vera, and B.~E. White, ``{Mass bounds in a model with a
  triplet Higgs},'' \href{http://dx.doi.org/10.1088/1126-6708/2003/06/059}{{\em
  JHEP} {\bfseries 06} (2003) 059},
  \href{http://arxiv.org/abs/hep-ph/0302256}{{\ttfamily arXiv:hep-ph/0302256}}.

\bibitem{Chen:2006pb}
M.-C. Chen, S.~Dawson, and T.~Krupovnickas, ``{Higgs triplets and limits from
  precision measurements},''
  \href{http://dx.doi.org/10.1103/PhysRevD.74.035001}{{\em Phys. Rev. D}
  {\bfseries 74} (2006) 035001},
  \href{http://arxiv.org/abs/hep-ph/0604102}{{\ttfamily arXiv:hep-ph/0604102}}.

\bibitem{Chivukula:2007koj}
R.~S. Chivukula, N.~D. Christensen, and E.~H. Simmons, ``{Low-energy effective
  theory, unitarity, and non-decoupling behavior in a model with heavy
  Higgs-triplet fields},''
  \href{http://dx.doi.org/10.1103/PhysRevD.77.035001}{{\em Phys. Rev. D}
  {\bfseries 77} (2008) 035001},
  \href{http://arxiv.org/abs/0712.0546}{{\ttfamily arXiv:0712.0546 [hep-ph]}}.

\bibitem{Bandyopadhyay:2020otm}
P.~Bandyopadhyay and A.~Costantini, ``{Obscure Higgs boson at Colliders},''
  \href{http://dx.doi.org/10.1103/PhysRevD.103.015025}{{\em Phys. Rev. D}
  {\bfseries 103} no.~1, (2021) 015025},
  \href{http://arxiv.org/abs/2010.02597}{{\ttfamily arXiv:2010.02597
  [hep-ph]}}.

\bibitem{Konetschny:1977bn}
W.~Konetschny and W.~Kummer, ``{Nonconservation of Total Lepton Number with
  Scalar Bosons},'' \href{http://dx.doi.org/10.1016/0370-2693(77)90407-5}{{\em
  Phys. Lett. B} {\bfseries 70} (1977) 433--435}.

\bibitem{Schechter:1980gr}
J.~Schechter and J.~W.~F. Valle, ``{Neutrino Masses in $SU(2) \times U(1)$
  Theories},'' \href{http://dx.doi.org/10.1103/PhysRevD.22.2227}{{\em Phys.
  Rev. D} {\bfseries 22} (1980) 2227}.

\bibitem{Cheng:1980qt}
T.~P. Cheng and L.-F. Li, ``{Neutrino Masses, Mixings and Oscillations in
  $SU(2) \times U(1)$ Models of Electroweak Interactions},''
  \href{http://dx.doi.org/10.1103/PhysRevD.22.2860}{{\em Phys. Rev. D}
  {\bfseries 22} (1980) 2860}.

\bibitem{Mohapatra:1980yp}
R.~N. Mohapatra and G.~Senjanovic, ``{Neutrino Masses and Mixings in Gauge
  Models with Spontaneous Parity Violation},''
  \href{http://dx.doi.org/10.1103/PhysRevD.23.165}{{\em Phys. Rev. D}
  {\bfseries 23} (1981) 165}.

\bibitem{Lazarides:1980nt}
G.~Lazarides, Q.~Shafi, and C.~Wetterich, ``{Proton Lifetime and Fermion Masses
  in an SO(10) Model},''
  \href{http://dx.doi.org/10.1016/0550-3213(81)90354-0}{{\em Nucl. Phys. B}
  {\bfseries 181} (1981) 287--300}.

\bibitem{Arhrib:2011uy}
A.~Arhrib, R.~Benbrik, M.~Chabab, G.~Moultaka, M.~C. Peyranere, L.~Rahili, and
  J.~Ramadan, ``{The Higgs Potential in the Type II Seesaw Model},''
  \href{http://dx.doi.org/10.1103/PhysRevD.84.095005}{{\em Phys. Rev. D}
  {\bfseries 84} (2011) 095005},
  \href{http://arxiv.org/abs/1105.1925}{{\ttfamily arXiv:1105.1925 [hep-ph]}}.

\bibitem{LHCHiggsCrossSectionWorkingGroup:2013rie}
{\bfseries LHC Higgs Cross Section Working Group} Collaboration, J.~R. Andersen
  {\em et~al.}, ``{Handbook of LHC Higgs Cross Sections: 3. Higgs
  Properties},'' \href{http://arxiv.org/abs/1307.1347}{{\ttfamily
  arXiv:1307.1347 [hep-ph]}}.

\bibitem{Chun:2012jw}
E.~J. Chun, H.~M. Lee, and P.~Sharma, ``{Vacuum Stability, Perturbativity, EWPD
  and Higgs-to-diphoton rate in Type II Seesaw Models},''
  \href{http://dx.doi.org/10.1007/JHEP11(2012)106}{{\em JHEP} {\bfseries 11}
  (2012) 106}, \href{http://arxiv.org/abs/1209.1303}{{\ttfamily arXiv:1209.1303
  [hep-ph]}}.

\bibitem{ATLAS:2020tlo}
{\bfseries ATLAS} Collaboration, G.~Aad {\em et~al.}, ``{Search for heavy
  resonances decaying into a pair of Z bosons in the $\ell ^+\ell ^-\ell
  '^+\ell '^-$ and $\ell ^+\ell ^-\nu {{\bar{\nu }}}$ final states using 139
  $\mathrm {fb}^{-1}$ of proton\textendash{}proton collisions at $\sqrt{s} =
  13\,$TeV with the ATLAS detector},''
  \href{http://dx.doi.org/10.1140/epjc/s10052-021-09013-y}{{\em Eur. Phys. J.
  C} {\bfseries 81} no.~4, (2021) 332},
  \href{http://arxiv.org/abs/2009.14791}{{\ttfamily arXiv:2009.14791
  [hep-ex]}}.

\bibitem{CMS:2024vps}
{\bfseries CMS} Collaboration, ``{Search for heavy scalar resonances decaying
  to a pair of Z bosons in the 4-lepton final state at 13 TeV},''.

\bibitem{ATLAS:2020jgy}
{\bfseries ATLAS} Collaboration, G.~Aad {\em et~al.}, ``{Search for the $HH
  \rightarrow b \bar{b} b \bar{b}$ process via vector-boson fusion production
  using proton-proton collisions at $\sqrt{s} = 13$ TeV with the ATLAS
  detector},'' \href{http://dx.doi.org/10.1007/JHEP07(2020)108}{{\em JHEP}
  {\bfseries 07} (2020) 108}, \href{http://arxiv.org/abs/2001.05178}{{\ttfamily
  arXiv:2001.05178 [hep-ex]}}. [Erratum: JHEP 01, 145 (2021), Erratum: JHEP 05,
  207 (2021)].

\bibitem{ATLAS:2022enb}
{\bfseries ATLAS} Collaboration, G.~Aad {\em et~al.}, ``{Search for heavy
  resonances decaying into a $Z$ or $W$ boson and a Higgs boson in final states
  with leptons and $b$-jets in $139~$fb$^{-1}$ of $pp$ collisions at
  $\sqrt{s}=13~$TeV with the ATLAS detector},''
  \href{http://dx.doi.org/10.1007/JHEP06(2023)016}{{\em JHEP} {\bfseries 06}
  (2023) 016}, \href{http://arxiv.org/abs/2207.00230}{{\ttfamily
  arXiv:2207.00230 [hep-ex]}}.

\bibitem{Butterworth:2023rnw}
J.~Butterworth, H.~Debnath, P.~Fileviez~Perez, and F.~Mitchell, ``{Custodial
  symmetry breaking and Higgs boson signatures at the LHC},''
  \href{http://dx.doi.org/10.1103/PhysRevD.109.095014}{{\em Phys. Rev. D}
  {\bfseries 109} no.~9, (2024) 095014},
  \href{http://arxiv.org/abs/2309.10027}{{\ttfamily arXiv:2309.10027
  [hep-ph]}}.

\bibitem{ATLAS:2022tnm}
{\bfseries ATLAS} Collaboration, G.~Aad {\em et~al.}, ``{Measurement of the
  properties of Higgs boson production at $\sqrt{s} = 13$ TeV in the
  $H\to\gamma\gamma$ channel using $139$ fb$^{-1}$ of $pp$ collision data with
  the ATLAS experiment},''
  \href{http://dx.doi.org/10.1007/JHEP07(2023)088}{{\em JHEP} {\bfseries 07}
  (2023) 088}, \href{http://arxiv.org/abs/2207.00348}{{\ttfamily
  arXiv:2207.00348 [hep-ex]}}.

\bibitem{CMS:2021kom}
{\bfseries CMS} Collaboration, A.~M. Sirunyan {\em et~al.}, ``{Measurements of
  Higgs boson production cross sections and couplings in the diphoton decay
  channel at $ \sqrt{\mathrm{s}} $ = 13 TeV},''
  \href{http://dx.doi.org/10.1007/JHEP07(2021)027}{{\em JHEP} {\bfseries 07}
  (2021) 027}, \href{http://arxiv.org/abs/2103.06956}{{\ttfamily
  arXiv:2103.06956 [hep-ex]}}.

\end{thebibliography}\endgroup

\end{document}